\documentclass[12pt,a4paper]{article}

\usepackage{amsmath,amssymb}
\usepackage[dvips]{graphicx}

\newcommand{\beq}{\begin{eqnarray}}
\newcommand{\eeq}{\end{eqnarray}}
\newcommand{\be}{\begin{equation}}
\newcommand{\ee}{\end{equation}}

\title{THEORY OF QUARK-GLUON PLASMA AND PHASE TRANSITION.}

\author{ E.V.Komarov \footnote{e-mail: bartnovsky@itep.ru}, Yu.A.Simonov \footnote{e-mail:
simonov@itep.ru}\\
State Research
Center\\Institute of Theoretical and Experimental Physics, \\
Moscow, 117218 Russia}

\begin{document}

\maketitle


\begin{abstract}
Nonperturbative picture of strong interacting
quark-gluon plasma is given based on the systematic Field
Correlator Method. Equation of state, phase transition in
density-temperature plane is derived and compared to lattice data
as well as subsequent thermodynamical quantities of QGP.
\end{abstract}

\section{Introduction}
The perturbative exploring of quark-gluon plasma (QGP) has some
difficulties in describing the physics of QGP and phase
transitions. However, it was realized 30 years ago that
nonperturbative (np) vacuum fields are strong (\cite{1}) and later
it was predicted (\cite{2}) and confirmed on the lattice
(\cite{3}) that the magnetic part of gluon condensate does not
decrease at $T>T_c$ and even grows as $T^4$ at large $T$ (\cite{3a}).

Therefore it is natural to apply the np approach, the Field
Correlator Method (FCM) (\cite{4}) to the problem of QGP and phase
transitions, which was done in a series of papers
(\cite{5}-\cite{9}). As a result one obtains np equation of state
(EoS) of QGP and the full picture of phase transition, including
an unbiased prediction for the critical temperature $T_c(\mu)$ for
different number of flavors $n_f$.

%\section{Brief derivation of EoS of QGP}

\section{Nonperturbative EoS of QGP} We split the gluonic
field $A_\mu$ into the background field $B_\mu$ and the (valence
gluon) quantum field $a_\mu$: $A_\mu=B_\mu+a_\mu$ both satisfying
the periodic boundary conditions.

The partition function averaged both in perturbative and np fields
is \be Z(V,T)=\langle Z(B+a)\rangle_{B,a}\label{1}\ee

Exploring free energy $F(T,\mu)=-T \ln \langle Z(B)\rangle_B$ that
contains perturbative and np interactions of quarks and gluons
(which also includes creation and dissociation of bound states) we
follow so-called Single Line Approximation (SLA). Namely, we
assume that quark-gluon system for $T>T_c$ stays gauge invariant,
as it was for $T<T_c$, and neglect all perturbative interactions
in the first approximation. Nevertheless in SLA already exist a
strong interaction of gluons (and quarks) with np vacuum fields.
This interaction consists of colorelectric (CE) and colormagnetic
(CM) parts. The CE part in deconfinement phase creates np
self-energy contribution for every quark and gluon embedded in
corresponding Polyakov line. An important point is that Polyakov
line is computed from the gauge invariant $q\bar q$ (gg) Wilson
loop, which for np $D_1^E$ interaction splits into individual
quark (gluon) contributions. As for CM part - its consideration is
beyond the SLA, because as has been recently shown in paper
(\cite{10})
 strong CM fields are responsible for creation of
bound states of white combinations of quarks and gluons.

To proceed with FCM we apply the nonabelian Stokes theorem and the
Gaussian approximation to compute the Polyakov line in terms of np field correlators

\begin{multline*}
L_{fund} = \frac{1}{N_c} tr\ P\exp\left(ig \int_0^\beta B_4(z)dz^4\right)= \\
\frac{1}{N_c} tr\
\exp\left(-\frac{g^2}{2}\int_{S_n}\int_{S_n}d\sigma_{\mu\nu}(u)
d\sigma_{\lambda\sigma}(v)D_{\mu\nu,\lambda\sigma}\right) %\label{}
\end{multline*}

with \be D_{\mu\nu,\lambda\sigma}\equiv g^2\langle
F_{\mu\nu}(u)\Phi(u,v)F_{\lambda\sigma}(v)\Phi(v,u)\rangle
\label{2} \ee

$D_1^E$ and $D^E$ arise from CE field strengths: \be \frac{1}{N_c}
D_{0i,0k}= \delta_{ik} \left[D^E+D_1^E+u_4^2 \frac{\partial D_1^E}
{\partial u_4^2} \right]+u_i u_k \frac{\partial D_1^E}{\partial
\vec{u}^2}\label{3}\ee

As a result the Polyakov loop can be expressed in terms of
"potentials" $V_1$ and $V_D$
\be
L_{fund}=\exp\left(-\frac{V_1(T)+2V_D}{2T}\right),
L_{adj}=\left(L_{fund}\right)^{9/4}, \label{4} \ee

with $V_1(T)\equiv V_1(\infty,T)$, $V_D\equiv V_D(r^*,T)$
(\cite{5}) \be V_1(r,T)=\int_0^\infty d\nu (1-\nu T) \int_0^r
d\xi\ \xi D_1^E(\sqrt{\xi^2+\nu^2})\label{5}\ee

\be V_D(r,T)=2\int_0^\infty d\nu (1-\nu T)\int_0^r d\xi\ (r-\xi)
D^E(\sqrt{\xi^2+\nu^2})\label{6}\ee

In what follows we use the Polyakov line fit (\cite{8,9}) \be
L_{fund}\left(x=\frac{T}{T_c},T\right)=\exp \left(
-\frac{.175\mbox{Gev}}{(1.35x-1)2T}    \right) \label{7} \ee

The free energy $F(T)$ of quarks and gluons in SLA can be
expressed as a sum over all Matsubara winding numbers $n$
 with coefficients $L_{fund}^n$ and $L_{adj}^n$ for quarks and
 gluons respectively. For nonzero chemical potential $\mu$ one can
  keep $L_{fund,adj}$ independent of $\mu$, treating np fields
 as strong and unchanged by $\mu$ in the first approximation.

The final formulas for pressure of qgp are (\cite{7,9})
 \be p_q\equiv
\frac{P^{SLA}_q}{T^4}=\frac{n_f}{\pi^2}\left[\Phi_\nu\left(\frac{\mu-\frac{V_1}{2}}{T}\right)+
\Phi_\nu\left(-\frac{\mu+\frac{V_1}{2}}{T}\right)\right]\label{8}\ee where
$\nu=m_q/T$ and
\be\Phi_\nu(a)=\int_0^\infty\frac{z^4}{\sqrt{z^2+\nu^2}}\frac{1}{(e^{\sqrt{z^2+\nu^2}}+1)}\label{9}\ee
\be p_{gl}=\frac{P^{SLA}_{gl}}{T^4}=\frac{8}{3\pi^2}\int_0^\infty
\frac{z^3dz}{e^{z+9/4 V_1}-1}\label{10}\ee

The energy density is $\varepsilon=T^2\frac{\partial}{\partial
T}\left(\frac{P}{T}\right)_V$ and the speed of sound in plasma is
$c_s^2=\frac{\partial P}{\partial \varepsilon}$. In Fig.5 $c_s^2$
is shown calculated with the use of (\ref{8}), (\ref{10}) and
compared to lattice data for $\mu=0$ from (\cite{11}). No lattice
calculations has yet been done for sound speed at nonzero baryon
density, though our theory allows to do that and as is shown in
(\cite{12}) the result for $\mu>0$ does not differ much from the
case $\mu=0$.

\begin{figure}[t]
  \begin{minipage}{5cm}
     \centering
     \includegraphics[scale=0.55]{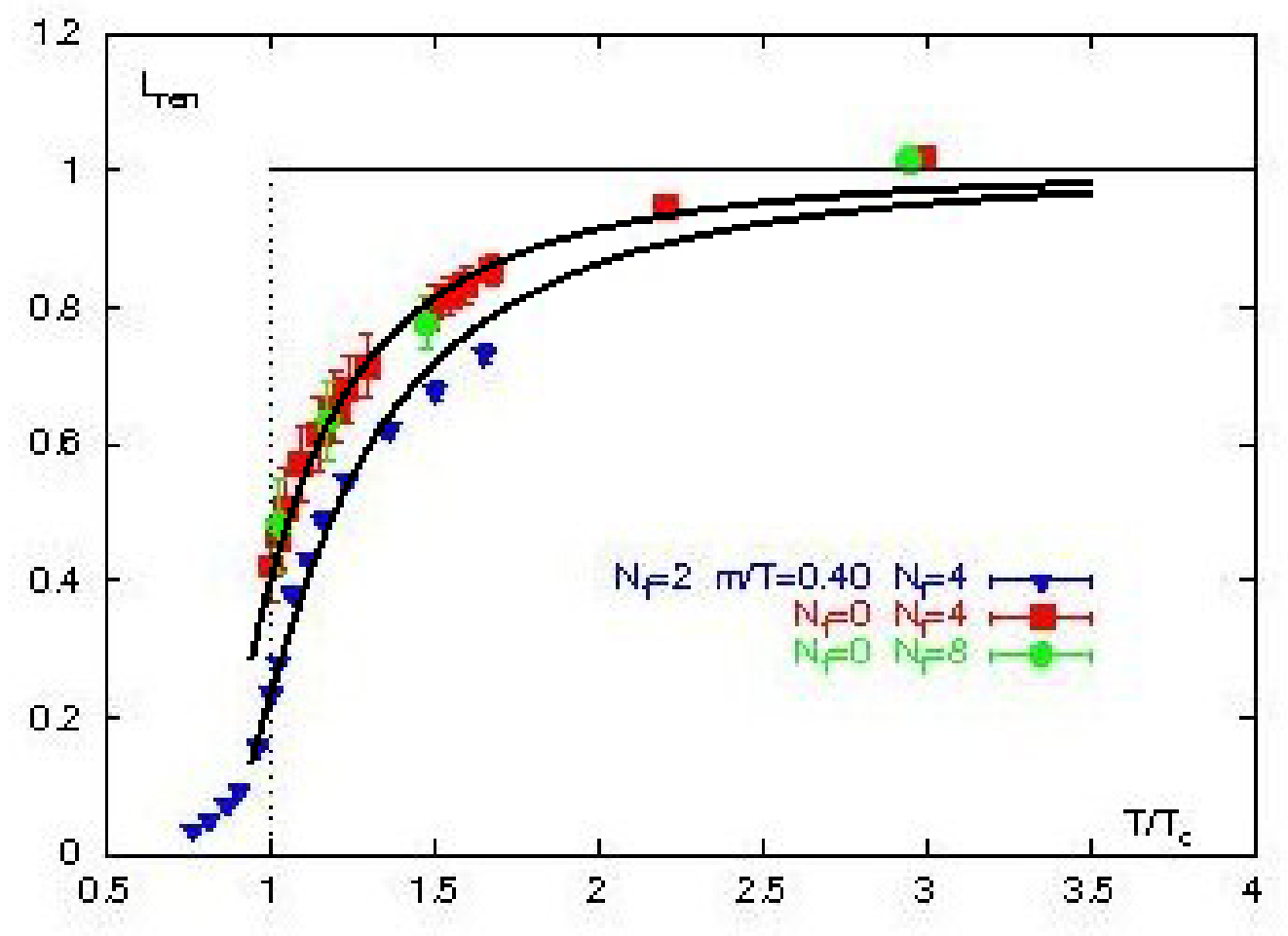}
     \caption{Fit (\ref{7}) of Polyakov line for $n_f=0$ and
     $n_f=2$)(black curves) to the lattice data (\cite{11}).}
     \label{}
  \end{minipage}
\hfill
  \begin{minipage}{5cm}
     \centering
     \includegraphics[scale=0.45]{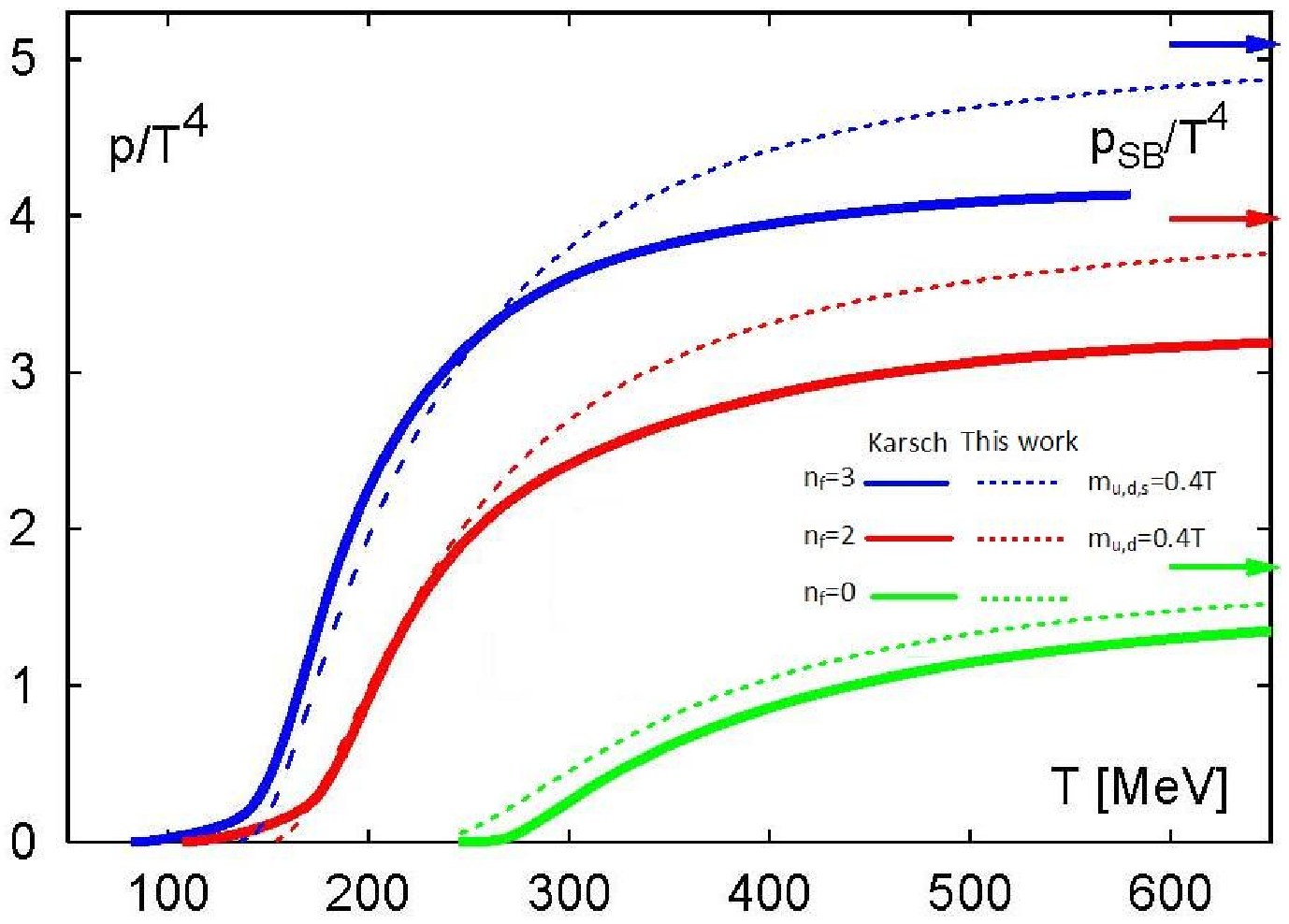}
     \caption{Analytic (\ref{8}), (\ref{10}) and lattice (\cite{11}) curves for pressure of QGP with $n_f=0,2+1,3$ from (\cite{9}).}
     \label{}
  \end{minipage}
\end{figure}

\begin{figure}[h]
  \begin{minipage}{5cm}
     \centering
     \includegraphics[scale=0.5]{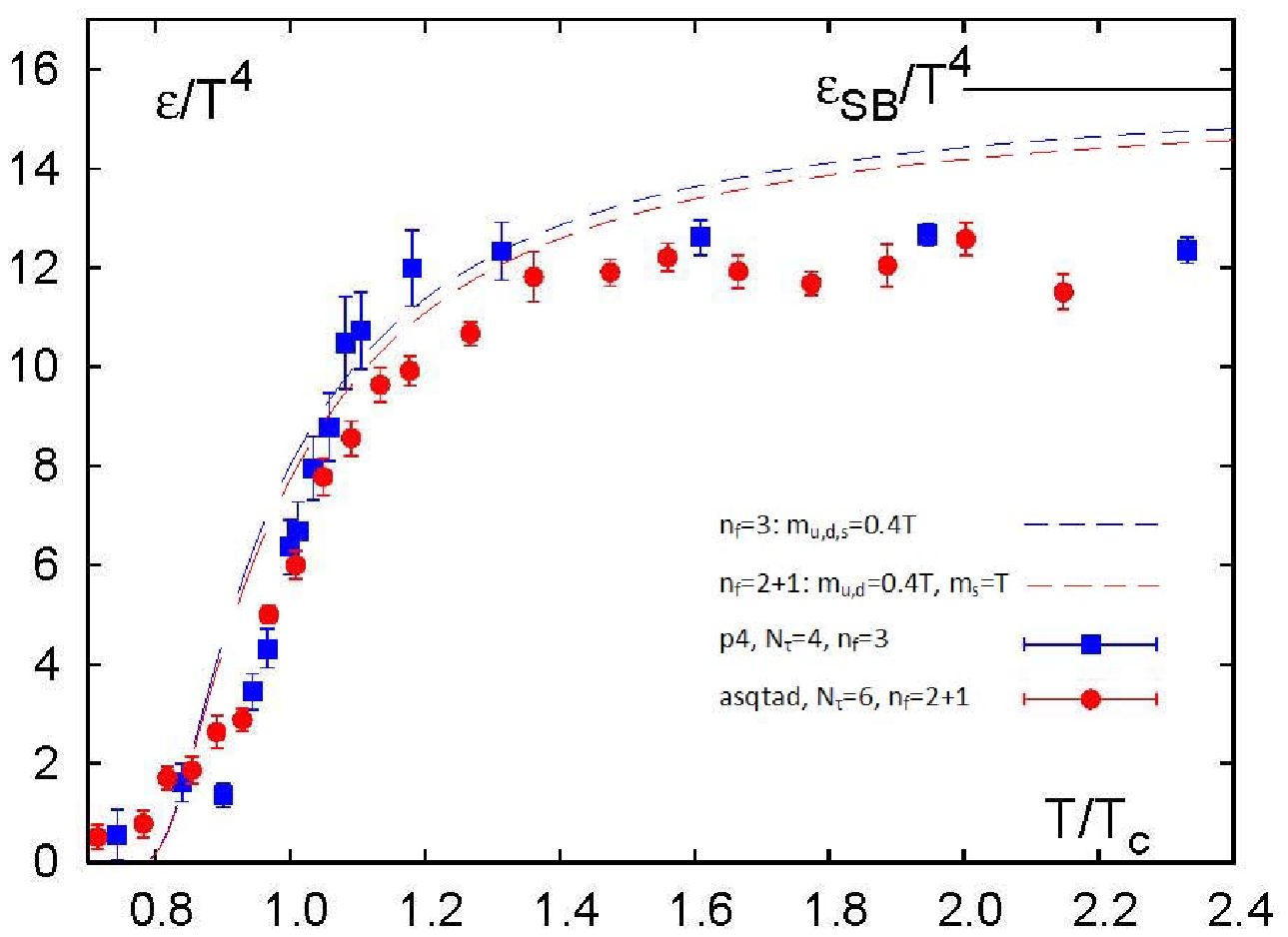}
     \caption{Analytic and lattice (\cite{11}) curves for energy density of QGP with $n_f=2+1$ and $n_f=3$ from (\cite{9}).}
     \label{}
  \end{minipage}
\hfill
  \begin{minipage}{5cm}
     \centering
     \includegraphics[scale=0.35]{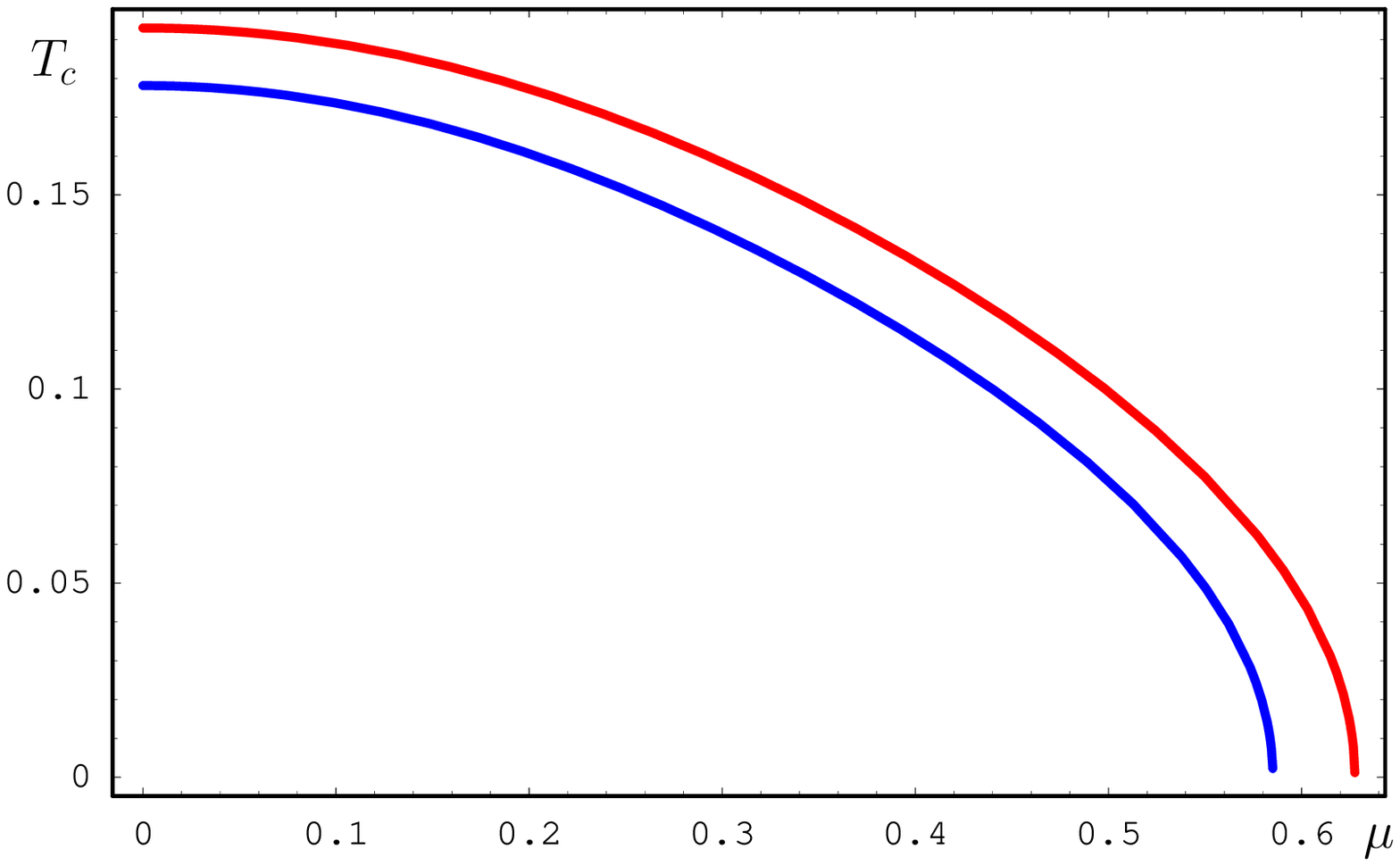}
     \caption{The phase transition curve $T_c(\mu)$ (in GeV) from
     (\ref{9}) as function of quark chemical potential $\mu$ (in GeV)
      for $n_f=2$ (upper curve) and $n_f=3$ (lower curve) and
      $\Delta G_2=0.0034\ \mbox{GeV}^4$ from (\cite{8}).}
     \label{}
  \end{minipage}
\end{figure}

\begin{figure}[!h]
\centering
\includegraphics[height=8cm]{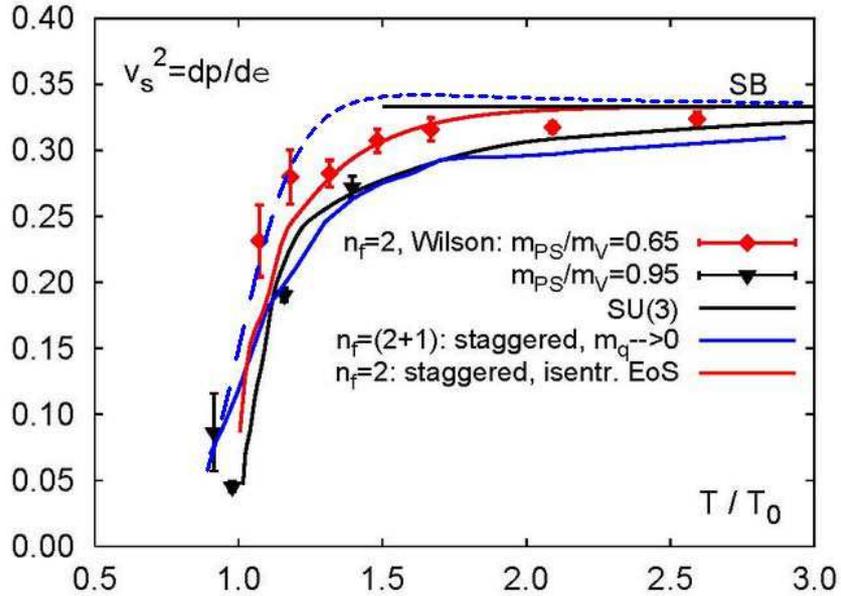}
\caption{Sound speed for $\mu=0$ and $n_f=3$ (blue dashed curve)
compared to lattice data from (\cite{11}).}
\end{figure}

\section{Phase transition} To obtain the curve of phase transition
one needs to define pressure $P_I$ in the confined phase and
$P_{II}$ in the deconfined phase, taking into account that vacuum
energy density contributes to the free energy, and hence to the
pressure: \be P_I=|\varepsilon_{vac}|+P_{hardon},\ \ \
P_{II}=|\varepsilon_{vac}^{dec}|+(p_q+p_{gl})T^4. \label{11}\ee
Having formulas for pressure (which contain parameter of
$L_{fund}(x)$ (\ref{7})) we may write for the phase
transition curve $T_c(\mu)$: \be
T_c(\mu)=\left(\frac{(11-\frac{2}{3}n_f)\Delta G_2}{32
(p_q+p_{gl})}\right)^{1/4}\label{12} \ee here $\Delta
\varepsilon_{vac}=|\varepsilon_{vac}-\varepsilon_{vac}^{dec}|=(11-\frac{2}{3}n_f)/32\Delta
G_2$. In particular, for the expected value of $\Delta G_2/G_2(stand)\approx 0.4$ one obtains
 $T_c=0.27 \mbox{ GeV } (n_f=0),\ 0.19 \mbox{ GeV } (n_f=2),\ 0.17 \mbox{ GeV } (n_f=3)$ in
 good agreement with lattice data.

\section{Summary} The EoS of QGP is written, where the
only np input is the Polyakov line. It should be stressed, that
only the modulus of the Polyakov line enters in EoS due to gauge
invariance. The phase transition curve $T_c(\mu)$ and speed of
sound $c_s^2(T)$ are obtained and agree well with lattice data. An
important point of the work is that the only parameter used to
receive the final physical quantities from the initial QCD
Lagrangian is the Polyakov line taken from lattice data, and is in
agreement with analytic estimate for $T=T_c$ (\cite{5}).

\section{Acknowledgments} The financial support of RFFI grant
06-02-17012 is acknowledged.

\end{document}